\documentclass[a4paper,11pt]{article}
\pdfoutput=1 % if your are submitting a pdflatex (i.e. if you have
\usepackage{jheppub} % for details on the use of the package, please
\usepackage{mathtools,slashed}

\def\be{\begin{equation}}
\def\ee{\end{equation}}
\def\beqn{\begin{eqnarray}}
\def\eeqn{\end{eqnarray}}

\def\ba{\begin{array}{c}}
\def\bat{\begin{array}{cc}}
\def\ea{\end{array}}
\def\bi{\begin{itemize}}
\def\ei{\end{itemize}}

\def\be{\begin{equation}}
\def\ee{\end{equation}}

\newcommand{\lag}{\mathcal{L}}

\definecolor{Darkgreen}{RGB}{30,120,30}

\def\bs{\begin{subequations}}
\def\es{\end{subequations}}

\renewcommand{\geq}{\geqslant}

\newcommand{\tia}[1]{}

\newcounter{listcounter}

\title{\boldmath Non-metricity signatures on the Higgs boson signal strengths at the LHC}

\author[a]{Victor Ilisie}
\emailAdd{victor.ilisie@universidadeuropea.es}

\affiliation[a]{Escuela de Ciencias, Ingeniería y Diseño, Universidad Europea de Valencia, \\ Paseo de la Alameda 7, 46010, Val\`encia, Spain}

\abstract{In this work we study the high-energy Higgs boson phenomenology associated to the non-metricity scale $\Lambda_Q$ at the LHC. Non-metricity is present in more generic non-Riemannian geometries describing gravity beyond General Relativity and exhibits nice features in astronomy and cosmology, and it can be analysed perturbatively. Using effective field theory tools, we calculate the {\it new physics} contributions to the one-loop $H\to\gamma\gamma$ and $gg\to H$ processes and, together with previous bounds from Compton scattering, we obtain relevant constraints and correlations in the model's parameter space. This can help us take a step further, and no longer associate gravitational effects uniquely to astronomical phenomena, and to start analysing these effects by means of high energy experiments. In turn, this could also help us get a better grasp at quantum phenomena associated to gravity.}

\begin{document}

\maketitle
\flushbottom

\section{Introduction}
\label{sec:introd}

With the discovery of the Higgs boson \cite{ATLAS:2012yve, CMS:2012qbp} the Standard Model (SM) has proven once more its success. Even though the latest provided data seem to be extraordinarily consistent with its predictions \cite{ATLAS:2022fnp,ATLAS:2022yrq,ATLAS:2021tbi,ATLAS:2021pkb,ATLAS:2020wny,
CERN-EP-2022-120,arXiv:2204.12945,arXiv:2204.12957,CMS:2020gsy,CMS:2018uag,CMS:2018nsn,CMS:2020dvg,CMS:2017dib,ATLAS:2021qou,Tumasyan:2801541, 
ATLAS:2020bhl,Sirunyan:2765059,Sirunyan:2753947,Sirunyan:2743741,ATLAS:2020cvh,ATLAS:2020jwz,ATLAS:2020fzp,ATLAS:2019nkf,Cadamuro:2019tcf,
ATLAS:2019aqa,ATLAS:2018mme,CMS:2022uhn} its discovery has left many open questions, such as, the need for additional CP violation in order to explain the matter-antimatter asymmetry in the Universe, the lack of candidates for dark matter and dark energy, the Higgs and quark hierarchy problem, just to name a few. As there is no fundamental principle that forbids us from extending this model, in the last few decades a large number of SM extensions have been proposed and analysed and, using the available experimental data, important bounds have been set on their parameter space. Nonetheless, if one does not wish to compromise with a specific extension, one can embed the SM Lagrangian into a generic effective field theory (EFT) Lagrangian as it has been extensively analysed in the literature \cite{Contino:2013kra,Bahl:2022yrs,Grzadkowski:2010es,Bechtle:2022tck,Kanemura:2021fvp,Battaglia:2021nys,Asiain:2021lch,Breso-Pla:2021qoe,Banerjee:2020vtm,DasBakshi:2020ejz,Hays:2020scx,Dobado:2019fxe,Sanz-Cillero:2020szj,Pich:2020xzo,Pich:2018ltt,Pich:1998xt} i.e., 
\begin{align}
    \mathcal{L}_{\text{EFT}} = \mathcal{L}_{\text{SM}} + \sum_{k,j}  \frac{c_j^{(k)}}{\Lambda^k} \, \mathcal{O}_j^{(k)} \, ,
\end{align}  
where $k\geq 1$ is the power corresponding to some new physics scale $\Lambda$ (given in mass units). The $c_j^{(k)}$ terms are the so-called Wilson coefficients and, $\mathcal{O}_j^{(k)}$ the operators built in terms of the SM fields, both corresponding to the $1/\Lambda^{k}$ term of the expansion. As it is already well known, the only drawback is that the previous expansion is infinite and therefore non-renormalizable. However, once the power series is {\it cut} at some power $k'$, the corresponding $\mathcal{L}_{\text{EFT}}$ Lagrangian is renormalizable in the usual sense \cite{Pich:1998xt}.

On the other hand, in a very similar situation we find General Relativity (GR). It has repeatedly proven its success over the past decades, and perhaps one of the most outstanding predictions, verified experimentally in the last years, is the existence of gravitational waves \cite{LIGOScientific:2016aoc,LIGOScientific:2016dsl,LIGOScientific:2016sjg,LIGOScientific:2017bnn,LIGOScientific:2017ycc,LIGOScientific:2017vox,LIGOScientific:2018mvr,LIGOScientific:2021djp,LIGOScientific:2020ibl}. Let us, however, comment on its theoretical structure and explain how, just as in the SM model case, it can be embedded in a larger and more generic framework. First of all, at the time GR was born, only Riemannian geometry was known, and so the model was built upon the metricity assumption i.e., 
\begin{align}
\nabla_\mu \, g_{\alpha\beta} = 0 \, ,
\end{align}
where $\nabla_\mu$ is the covariant derivative and $g_{\alpha\beta}$ the metric components. However, nowadays we know that, for a generic manifold, given a metric $g$ and a connection $\Gamma$, we can define the non-metricity (Q), curvature (R) and torsion (S) tensors as three independent quantities, as it is nicely described in \cite{Mao:2006bb}
\begin{align}
Q_{\mu\nu\rho} &= \nabla_\nu \, g_{\nu\rho} \, , \notag  \\
R^\rho_{\;\; \lambda\nu\mu} &= \Gamma^\rho_{\;\; \mu\lambda,\nu} - \Gamma^\rho_{\;\; \nu\lambda,\mu} + \Gamma^\rho_{\;\; \nu\alpha}\Gamma^\alpha_{\;\; \mu\lambda} - \Gamma^\rho_{\;\; \mu\alpha}\Gamma^\alpha_{\;\; \nu\lambda} \,, \\
S_{\mu\nu}^{\;\;\;\;\rho} &= \frac{1}{2} \left( \Gamma^\rho_{\;\;\mu\nu} -  \Gamma^\rho_{\;\;\nu\mu} \right)  \, , 
\end{align}
where $\Gamma^\rho_{\;\; \mu\lambda,\nu} \equiv \partial_\nu \, \Gamma^\rho_{\;\; \mu\lambda}$, with $\partial_\nu$ the ordinary partial derivative and $\Gamma^{\mu}_{\;\;\nu\alpha}$ the components of the connection $\Gamma$ in a given basis. Note therefore, that GR in its original form can be seen as described by a particular spacetime geometry with $Q = S = 0$ and, as a consequence, the connection $\Gamma$ can be written in terms of the metric tensor and its first (ordinary) derivatives.

Thus, again, just as in the case of the SM, as there is no fundamental principle that forbids us to extend GR, we can embed it into more generic geometrical theories that allow for non-vanishing torsion and non-metricity, and constrain these terms experimentally, as it is currently done in high energy particle physics. Depending on the specific characteristics of $S$ and $Q$, many models have been discussed and analyzed in the literature, such as \cite{Hammond:2002rm,Gronwald:1995em,DeAndrade:2000sf,RevModPhys.48.393,Watanabe:2004nt,Capozziello:2001mq,PhysRevLett.56.2873,Shapiro:2001rz,Baekler:2006de,PhysRevD.67.108501,Saa:1993fx,Hehl:1994ue,Delhom:2020hkb,Silva:2022pfd,Afonso:2021aho,Saridakis:2021vue,Dombriz:2021bnl,BeltranJimenez:2019odq,BeltranJimenez:2018vdo,BeltranJimenez:2019esp,BeltranJimenez:2017tkd,Jimenez-Cano:2020chm,Jimenez-Cano:2020lea,Jimenez-Cano:2022sds,Olmo:2022ops,deCesare:2016mml}, just to mention a few. 

Here, we are going to focus on Ricci-based gravity (RBG) models \cite{Afonso:2018bpv} and their high energy phenomenology at the LHC. These models are strongly motivated from a physical point of view \cite{Afonso:2018bpv,BeltranJimenez:2017doy,Pani:2012qb,Olmo:2022rhf,Delhom:2019wir,Latorre:2017uve,Ellis:2017edi,BeltranJimenez:2021oaq} and also, they result appealing as they can be treated perturbatively within an effective field theory framework \cite{BeltranJimenez:2021oaq}. Namely, RBGs only propagate spin-2 polarizations as in GR, and also, the corresponding gravitational waves, in vacuum, travel at the speed of light, which makes them compatible with the current experimental data, as shown in \cite{BeltranJimenez:2017doy,BeltranJimenez:2017uwv}. 
Moreover, RBGs are a class of metric-affine theories for which the field equations for the
metric are always of second order, which alleviates in many cases the problem associated
to the propagation of ghostly degrees of freedom \cite{BeltranJimenez:2019acz}. 
More interestingly, they have gained more attention in the past decade, as they present similar singularity-free solutions to Big-Bang cosmologies and black holes as some approaches to quantum gravity, therefore it has been suggested that they could be understood as a low-energy limit of a possible
quantum theory of gravity \cite{Olmo:2015bha, Lobo:2014nwa, Hossenfelder:2017rub, Olmo:2008nf}. All RBG theories present a branch of solutions where the gravitational dynamics is reproduced by Einstein’s equations (thus mimicking GR) coupled to some stress-energy tensor, however, some particular cases, such as Eddington-inspired Born-Infeld gravity (that we shall briefly comment upon in the following section) present specific couplings to matter which avoid
the appearance of singularities in early cosmology \cite{Afonso:2018bpv,BeltranJimenez:2017doy,Delhom:2019zrb}. However, it has been shown that quantum corrections in the non-perturbative regime might spoil the nice properties of these models \cite{BeltranJimenez:2021oaq,BeltranJimenez:2017uwv}.

In this paper, using the techniques introduced in \cite{Latorre:2017uve,Delhom:2019wir,BeltranJimenez:2021oaq,Delhom:2021bvq}, we are going to briefly introduce the RBG action and its expansion in powers of the non-metricity scale, and use it to calculate the effective Lagrangian corresponding to the $H\bar{f}f\gamma\gamma$ and $\bar{f}f\gamma\gamma$ vertices, where $f$ is an arbitrary fermion (with or without electric charge), $H$ the Higgs boson and $\gamma$, a photon. As we shall see, our results do not agree with the ones obtained in \cite{Delhom:2019wir} for the $\bar{f}f\gamma\gamma$ Lagrangian nor with the expression of the $\sigma_{\gamma e \to \gamma e}$ cross section including non-metricity effects. Therefore, we shall indicate, in each case, the differences that we find with the previously mentioned work and also, re-perform the phenomenological analysis presented therein in terms of the correct expression for the $\sigma_{\gamma e \to \gamma e}$ cross section. Afterwards, we shall use the same Lagrangian (corresponding to the $H\bar{f}f\gamma\gamma$ and $\bar{f}f\gamma\gamma$ vertices) for calculating the relevant Higgs decay and productions channels and the corresponding LHC signal strengths, again including modifications induced by the non-metricity scale. Finally, we will perform a phenomenological analysis comparing the model predictions with the currently available experimental data and extract bounds on the parameters of the model.

\section{Ricci-based Gravity Theories and EFTs}

As it was recently pointed out \cite{Latorre:2017uve,Delhom:2019wir,Olmo:2022rhf,Afonso:2018bpv}, there is a broad class of higher-order curvature theories that are characterized by non-trivial non-metricity tensors. In these studies, for a broad subset of the previously mentioned theories i.e., Ricci-based gravity theories, where non-metricity is sourced by local energy-momentum densities that cannot be gauged away by a projective transformation, it was found that the presence of non-metricity can induce perturbative effective interactions for fields with spin 0, 1/2 and 1. Within this class, there is a sub-class formed by projective-invariant RBG theories, that only include the symmetric part of the Ricci tensor.\footnote{It has been shown that, introducing the antisymmetric part of the Ricci tensor, generates ghostly degrees of freedom which present additional physical complications\cite{PhysRevD.33.2756,JOHNSTON1988721,Woodard:2015zca,BeltranJimenez:2019acz,BeltranJimenez:2020sqf} and we shall not consider such cases in our analysis.} The action, for this last sub-class of RBG theories, can be generically written as
\begin{align}
S_\text{RBG} = \frac{1}{2\kappa} \int d^4x \sqrt{-g} F_{\text{RBG}}[g_{\mu\nu}, R_{(\mu\nu)}, \Lambda_{\text{RBG}}] + S_\text{M}[g_{\mu\nu},\Psi, \Gamma^{\alpha}_{\;\;\mu\nu}] \, ,
\label{action}
\end{align}
where $F_{\text{RBG}}$ is an analytic function of $g_{\mu\nu}$, $R_{(\mu\nu)}$ and $\Lambda_{\text{RBG}}$, with $\kappa = M_{\text{Pl}}^{-1}$ and with $M_{\text{Pl}}$ Plank's mass. The $\Lambda_{\text{RBG}}$ factor is the scale at which deviations form GR come into the game, $R_{(\mu\nu)}$ is the symmetric part of the Ricci tensor and $S_\text{M}$, the action containing the matter fields $\Psi$ ($\Psi$ stands for a generic matter field of arbitrary spin and mass). 

As we shall see in the following, the metric in RBG theories, in general, can be expanded as
\begin{align}
g^{\mu\nu} = q^{\mu\nu} + \frac{1}{\Lambda_Q^4} \left( \alpha \, T \, q^{\mu\nu} + \beta \, T^{\mu\nu} \right) + \mathcal{O}(\Lambda_Q^{-8}) \, ,
\label{gmunuexp}
\end{align}
where $\alpha$ and $\beta$ are arbitrary parameters to be constrained experimentally, as well as $\Lambda_Q$. The quantities $T^{\mu \nu}$ and $T$ are the stress-energy tensor and its trace, and $q$ is the Einstein frame metric, i.e., the metric that satisfies the equations of GR coupled to a nonlinearly modified matter sector. In vacuum $q = g$, and GR dynamics is exactly recovered (at least for a branch of the RBG solutions \cite{BeltranJimenez:2020guo}). However, in the presence of matter, the deviations between $q$ and $g$ are suppressed by the scale $\Lambda_Q$, given by $\Lambda_Q = \sqrt{M_{\text{Pl}} \, \Lambda_{\text{RBG}}}$. Because the equations of RBG theories state that the connection is given by the Levi-Civita connection of $q$, it turns out that the non-metricity of $g$ is completely determined by these $\Lambda_Q$-suppressed corrections. Therefore, $\Lambda_Q$ can be regarded as both the non-metricity scale and the scale at which deviations from GR become non-perturbative, as explained in \cite{Latorre:2017uve,Delhom:2019wir}.

It has been proven\cite{Latorre:2017uve,Delhom:2019wir,Olmo:2022rhf,Afonso:2018bpv} that the action $S$ admits a representation, the Einstein frame representation (as mentioned previously), where the gravitational sector is described by standard GR for a metric $q_{\mu\nu}$ and where, the matter sector is minimally coupled to gravity.\footnote{The definition of {\it minimally coupled} in this context can be found in \cite{Delhom:2020hkb}.} In general, we can relate the metric tensor $g_{\mu\nu}$ corresponding to an arbitrary frame to the Einstein frame metric tensor, through the so-called deformation matrix $\Omega$ given by
\begin{align}
q_{\mu\alpha} \; (\Omega^{-1})^\alpha_{\;\;\nu}  = g_{\mu\nu}.
\end{align}
The deformation matrix is an on-shell function of the stress-energy tensor and can be expanded in powers of $1/\Lambda_Q^4$, similar to \eqref{gmunuexp}, as
\begin{align}
(\Omega^{-1})^\alpha_{\;\;\nu} = \delta^\alpha_{\;\;\nu} + \frac{1}{\Lambda_Q^4} \left( \alpha \,  T \, \delta^\alpha_{\;\;\nu}   + \beta \, T^\alpha_{\;\; \nu}\right) + \mathcal{O}(\Lambda_Q^{-8}) \, ,
\end{align}
where the first $\delta^\alpha_{\;\;\nu}$ term guarantees that GR is recovered as the low energy limit of the RBG model. It is worth mentioning that this expansion does not cover all the possible solutions, however, as argued in \cite{BeltranJimenez:2020guo}, the solutions that are not covered by this expansion normally suffer from physical pathologies.

We thus obtain, including terms of $\mathcal{O}(\Lambda_Q^{-4})$
\begin{align}
g_{\mu\nu} = q_{\mu\nu} + \frac{1}{\Lambda_Q^4} \left( \alpha \,  T \, \delta^\alpha_{\;\;\nu}   + \beta \, T^\alpha_{\;\; \nu}\right) \, .
\end{align}
Furthermore, for weak gravitational fields we can expand $q_{\mu\nu}$ about the Minkowskian metric simply as\footnote{This expansion is possible due to the fact that $q$ is the Einstein frame metric that satisfies the Einstein equations coupled to a non-linearly modified
matter sector. As the amount of matter (density) is small in our case, this will still be a
small correction from the vacuum solution to Einstein’s equations. Since
Minkowski is a vacuum solution of Einstein’s equations, the deviations
of $q$ from $\eta$ will be small.}
\begin{align}
q_{\mu\nu} = \eta_{\mu\nu} + \delta q_{\mu\nu},  
\end{align}
where $\delta q_{\mu\nu}$ encodes the Newtonian and post-Newtonian corrections to the metric (long range gravitational effects) that can be ignored for high-energy experiments on the Earth's surface.

In the previously introduced generic action (\ref{action}) we have not specified explicitly the functions $\text{F}_\text{RBG}$ nor the $\text{S}_\text{M}$ content. In this sense, particular cases of RBG models, that are worth mentioning, are the Eddington-inspired-Born-Infeld (EiBI) model (see  \cite{BeltranJimenez:2017doy} for a review) and quadratic $f(R)$ or $f(R, R_{(\mu\nu)}R^{(\mu\nu)})$ Palatini models \cite{Olmo:2012nx,Feola:2019zqg,Gialamas:2019nly,Sotiriou:2009xt,Olmo:2011uz,Afonso:2017bxr}.

The EiBI action is given by
\begin{align}
S_{\text{EiBI}} = \pm \Lambda_Q^4 \int d^4 x \left( \sqrt{-\left|g_{\mu\nu} \pm \Lambda_{\text{EiBI}}^{-2} R_{(\mu\nu)}(\Gamma) \right|}  -   \lambda \, \sqrt{-\left| g_{\mu\nu} \right|}      \right) \, ,
\end{align}
and it combines Eddington affine gravity with ideas from Born-Infeld (BI) electromagnetism,\footnote{The Born-Infeld (BI) \cite{Erskine:1935vmf,Ellis:2017edi,BeltranJimenez:2017doy,Afonso:2018bpv} modification of the QED action is given by
\begin{align}
S_{\text{BI}} = \beta^2 \int d^4 x \left( 1 - \sqrt{ 1 + 2\beta^{-2} F_{\mu\nu}F^{\mu\nu}  - 16\beta^{-4} \left(F_{\mu\nu}\tilde{F}^{\mu\nu}\right)^2 }\right) \, ,
\end{align}
and it was proposed to avoid the divergent self-energy problem of point charges in classical field theory. Interestingly, BI electrodynamics
coupled to GR is equivalent to EiBI gravity coupled to Maxwell electrodynamics, as shown in \cite{Delhom:2019zrb}.} where $\beta = \pm 1$ is the sign in front of $\Lambda_Q$. In this last case $\Lambda_Q$ is related to $\Lambda_{\text{EiBI}}$ by $\Lambda_Q = \sqrt{M_{\text{Pl}} \, \Lambda_{\text{EiBI}}}$. EiBI gravity exhibits nice features as it can yield non-singular solutions for different scenarios in cosmology and astrophysics. 

As for the $f(R)$ theories, the $F_{\text{RBG}}$ function from \eqref{action} simply has the particular form $F_{\text{RBG}}=f(R)$ where $R = g^{\mu\nu} R_{\mu\nu} = g^{\mu\nu} R_{(\mu\nu)}$ as $g^{\mu\nu}$ is always symmetric. Also, this case corresponds $\beta = 0$ \cite{Afonso:2017bxr}, which can be easily inferred from the fact that the deformation matrix is, in general, proportional to the identity matrix i.e., in the form
\begin{align}
\Omega^\mu_{\;\; \nu} = \frac{\partial f}{\partial R} \delta^\mu_{\;\; \nu} \, .
\end{align}

In this analysis we will extract and reinterpret the generic bounds obtained in our study, also in terms of the EiBI and the $f(R)$ models. However, one should be careful with the interpretation of the bounds on $f(R)$ models, as they are physically equivalent to metric-compatible models (with a trivial metricity tensor) with non-trivial torsion [108], therefore the {\it new physics} effects are not necessarily originated by genuine non-metricity phenomena.

\section{The effective $(H)\bar{f}f\gamma\gamma$ vertices}

Using the previous results, following similar procedures as in \cite{Latorre:2017uve,Delhom:2019wir}, we will deduce the effective Lagrangian corresponding to the $H\bar{f}f\gamma\gamma$ and the $\bar{f}f\gamma\gamma$
vertices, that will be used for recalculating the modified Compton Cross section $\sigma_{\gamma e \to \gamma e}$ and for calculating the Higgs signal strengths at the LHC.

Let us start by considering the following Lagrangian that corresponds to a massive Dirac fermion (which can be either charged or neutral), the kinetic term for the photon field, and the Yukawa interaction Lagrangian with all its terms in a curved space-time background i.e., with a generic metric $g^{\mu\nu}$:
\begin{align}
\mathcal{L}_{{\text{eff}}} &= \sqrt{-g} \bigg[ \frac{i}{2} e_a^{\;\;\mu} \bigg( \bar{\psi} \gamma^a \nabla_\mu\psi - (\nabla_\mu \bar{\psi}) \gamma^a \psi  \bigg) - m \bar{\psi}\psi  \notag \\
&\qquad\qquad\qquad\qquad +  \frac{1}{4} g^{\mu\nu}g^{\alpha\beta} F_{\mu\alpha} F_{\nu\beta} - \frac{m}{v} H \bar{\psi}\psi \bigg]     
 \, ,
\label{ini_lagr}
\end{align}
where $\nabla_\mu = \partial_\mu - \Gamma_\mu - B_\mu$, with $\Gamma_\mu$ the spinor connection 
\begin{align}
\Gamma_\mu =  \frac{i}{2} \omega_\mu^{\;\; ab} \frac{\sigma_{ab}}{2} \, , \qquad
\sigma_{ab} = \frac{i}{2} (\gamma_{a}\gamma_b - \gamma_{b}\gamma_{a}) \, ,
\end{align}
and where $B_\mu$ stands for the contributions of arbitrary gauge fields. As usual, the tetrads are given by $g^{\mu\nu} = e_a^{\;\;\mu} e_b^{\;\;\nu} \eta^{ab}$ with $\eta^{ab}$ the flat Minkowski metric. Expanding the tetrads $e_a^{\;\;\mu}$ up to $\mathcal{O}(\Lambda_Q^{-4})$, neglecting suppressed torsion effects \cite{Latorre:2017uve,Delhom:2019wir} induced by $\Gamma_\mu$, and also Newtonian and post-Newtonian corrections to the metric, we obtain 
\begin{align}
e_a^{\;\;\mu} = \delta_a^{\;\;\mu} - \frac{1}{2\Lambda_Q^4} (\alpha \,  T^{(J)}  \delta_a^{\;\;\mu} + \beta \, T_a^{(J)\mu}) \, .
\label{a1}
\end{align}
Therefore, the expression for the metric including perturbation terms reads
\begin{align}
g^{\mu\nu} = \eta^{\mu\nu} - \frac{1}{\Lambda^4_Q} (\alpha \, T^{(J)} \eta^{\mu\nu} + \beta T^{(J)\mu\nu} ) \, , 
\label{a2}
\end{align}
where $T_a^{(J)\mu}$ (or equivalently $T^{(J)\mu\nu}$) and $T^{(J)}$ are the stress-energy tensor and its trace for a spin-$J$ field. 

In this analysis, besides tree-level processes we are also interested in calculating loop-induced ones. The results in this last case will be UV-divergent and will, therefore, need renormalization. In order to regularize these divergences we will work in $\mathcal{D}=4+2\epsilon$ dimensions with $\epsilon < 0$, $|\epsilon|\ll 1$. To insure that we do not miss out possible finite contributions, as the previous terms of the interaction Lagrangian contain metric contractions, we will work in $\mathcal{D}$ dimensions already at the Lagrangian level. Using $\eta_{\mu\nu} \, \eta^{\mu\nu} = \mathcal{D}$ the expansion for $\sqrt{-g}$, up to $\mathcal{O}(\Lambda^{-4}_Q)$, reads
\begin{align}
 \sqrt{-g} = \sqrt{-\eta} \left( 1 + \frac{1}{2} \, \frac{\alpha \, T^{(J)} \eta^{\mu}_\mu + \beta \, T^{(J)\mu}_\mu}{\Lambda_Q^4} \right) = 1 + \frac{\mathcal{D}\alpha + \beta}{2\Lambda_Q^4} T^{(J)} \, .
\label{a3}
\end{align}
Inserting the expressions (\ref{a1}-\ref{a3}) into (\ref{ini_lagr}) and keeping terms up to $\mathcal{O}(\Lambda^{-4}_Q)$ we get 
\begin{align}
\mathcal{L}_{\text{eff}} &= \frac{i}{2}   \bar{\psi} \overleftrightarrow{\slashed{D}}\psi  - m \bar{\psi}\psi  + \frac{1}{4} F^{\mu\nu} F_{\mu\nu}  - \frac{m}{v} H \bar{\psi}\psi + \mathcal{L}_{(1/2)}^{\text{eff}} + \mathcal{L}_{(1)}^{\text{eff}} + \mathcal{L}_{(0)}^{\text{eff}} \, , 
\end{align}
where $D_\mu = \partial_\mu - B_\mu$, $\slashed{D}=\gamma^\mu D_\mu$ and $\bar{\psi} \overleftrightarrow{\slashed{D}}\psi = \bar{\psi} \gamma^\mu D_\mu \psi -  (D_\mu \bar{\psi})\gamma^\mu \psi$. 

Taking $J=1$ for the expansion of the tetrads and the $\sqrt{-g}$ term for the Dirac Lagrangian, $J=1/2$ for the expansion of the metric and the $\sqrt{-g}$ term for the photon field Lagrangian, and finally, $J=1$ for the expansion of $\sqrt{-g}$ for the Yukawa term, we obtain in flat Minkowsky space-time the complete set of interactions corresponding to the $H\bar{f}f\gamma\gamma$ and $\bar{f}f\gamma\gamma$ vertices. Their expressions are given by the sum of the three effective interaction Lagrangians $ \lag^{\text{eff}}_{Q} = \mathcal{L}_{(1/2)}^{\text{eff}}+\mathcal{L}_{(1)}^{\text{eff}}+\mathcal{L}_{(0)}^{\text{eff}}$ (see Appendix~\ref{INT_LAGR})\footnote{We would like to point out an erratum in \cite{Delhom:2019wir} i.e., for the effective photon interaction Lagrangian $\mathcal{L}_{s=1}^Q$. Our expression corresponds to $\mathcal{L}_{(1)}^{\text{eff}}$ from Appendix~\ref{INT_LAGR}, for the particular case $J=1/2$. On the other hand, we do agree with the expression for the effective four-photon vertex in $\mathcal{L}_{AA}^Q$ which can be obtained from our expression of $\mathcal{L}_{(1)}^{\text{eff}}$ with the substitution $T^{(1/2)}\to T^{(1)}$ and $T^{(1/2)\mu\nu}\to T^{(1)\mu\nu}$ from the same appendix. We can also observe that our result for the $\bar{f}f\gamma\gamma$ vertex (first term from \eqref{int_lagr}) differs in a factor 3 with respect to $\mathcal{L}_{\psi A}^Q$ from the same referred work. Also, as mentioned previously, we shall see that our result does not agree with the expression for the Compton cross section.} that explicitly reads
\begin{align}
 \lag^{\text{eff}}_{Q} &=   \frac{3 \, i \, \beta}{4\Lambda_Q^4} \, F^{\mu\alpha}F^{\nu}_{\;\;\alpha} \, \Big( \bar{\psi} \gamma_\mu \partial_\nu \psi - (\partial_\nu \bar{\psi})\gamma_\mu \psi \Big) - \left( \frac{\beta}{2} + \alpha \, \epsilon  \right) \frac{3}{4 \Lambda_Q^4} m \bar{\psi}  \psi \, F_{\alpha\beta} F^{\alpha\beta} 
\notag \\
 & -  \epsilon \left( 4 \alpha  + \beta  \right) \frac{m}{v} \frac{1}{4 \Lambda_Q^4} H \bar{\psi}  \psi \, F_{\alpha\beta} F^{\alpha\beta}
 \, .
\label{int_lagr}
\end{align}
One should note that, having extended the Lagrangian to $\mathcal{D}$ dimensions by using dimensional regularization, we will obtain additional contributions proportional to both $\alpha$ and $\beta$ in the considered loop diagrams.

\begin{figure}[!t]
\centering
\includegraphics[scale=0.55]{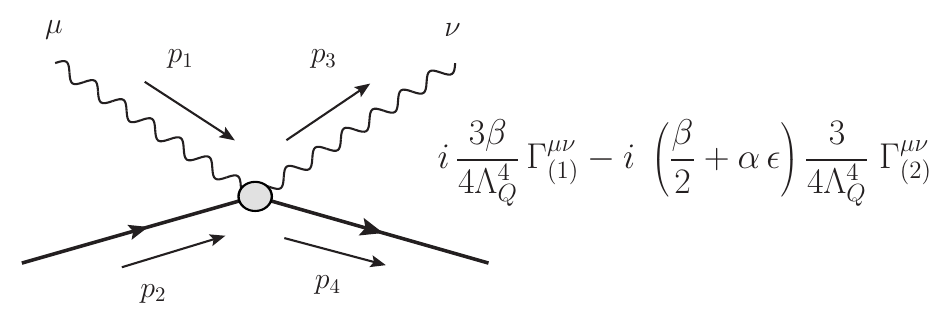}
\caption{Feynman rule corresponding to the $\bar{f}f\gamma\gamma$ effective interaction vertex and the corresponding four-momentum configuration.}
\label{eff_v}
\end{figure}

The corresponding Feynman rule for the new $\bar{f}f\gamma\gamma$ interaction term is shown in figure~\ref{eff_v}, where $\Gamma^{\mu\nu}_{(1)}$ and $\Gamma^{\mu\nu}_{(2)}$ explicitly read
\begin{align}
\Gamma^{\mu\nu}_{(1)} &=  \eta^{\mu\nu} \, A  - \slashed{p_1} \, p_3^\mu (p_2^\nu + p_4^\nu)  - \slashed{p_3} \, p_1^\nu (p_2^\mu + p_4^\mu) - p_1^\nu \, \gamma^\mu B
 \notag \\ 
                     &\qquad  - p_3^\mu \, \gamma^\nu C +   \gamma^\mu (p_2^\nu + p_4^\nu) (p_1 \cdot p_3)  +   \gamma^\nu (p_2^\mu + p_4^\mu) (p_1 \cdot p_3)      \, , \notag \\[1.5ex]
\Gamma^{\mu\nu}_{(2)} &=  4 m \, (\eta^{\mu\nu} \, p_1\cdot p_3  - \, p_3^\mu p_1^\nu ) \, ,                   
\end{align}
with the form factors $A,B$ and $C$ given by
\begin{align}
A &=  \slashed{p_1} \, B   + \slashed{p_3} \, C \,, 
\notag \\
B &=  \left( p_2 \cdot p_3  + p_3 \cdot p_4 \right) \, , 
\notag\\
C &= \left( p_1 \cdot p_2  + p_1 \cdot p_4 \right) \, .
\end{align}
One can check that this vertex, satisfies the corresponding Ward identities i.e., 
\begin{align}
p_{1,\mu} \, \Gamma^{\mu\nu}_{(1,2)} \, = 0  \, = \, p_{3,\nu} \, \Gamma^{\mu\nu}_{(1,2)} \, ,
\label{Wrd}
\end{align} 
for both on-shell and off-shell photons. The Feynman rule for the remaining interaction term, involving a Higgs boson, can be trivially obtained from the expression of $\Gamma^{\mu\nu}_{(2)}$.

\begin{figure}[!t]
\centering
\includegraphics[scale=0.4]{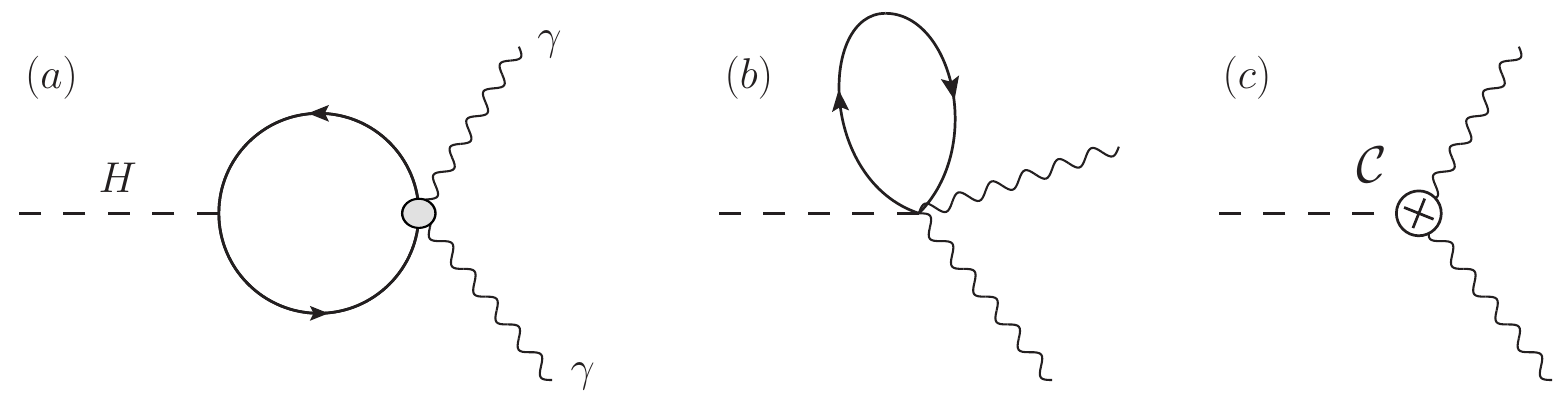} 
\caption{Contributing diagrams to the $H\to\gamma\gamma$ decay from the $\bar{f}f\gamma\gamma$ and $H\bar{f}f\gamma\gamma$ effective interaction vertices and the corresponding renormalization counterterm.}
\label{eff_mu}
\end{figure}

In this analysis we are interested in calculating the $H\to \gamma\gamma$ one loop-induced decay (figure~\ref{eff_mu}) and the $gg\to H$ production cross section. As the previous model is non-renormalizable we have to introduce a counterterm Lagrangian in order to re-absorb the UV divergences generated at the one-loop level for the $H\to\gamma\gamma$ process.\footnote{This proceedure can be trivially expanded to the $gg\to H$ process, but it will not be necessary as one can easily infer the finite expression for this process from the finite loop expression of $H\to \gamma\gamma$.} We shall parametrize it as
\begin{align}
\mathcal{L}_{Q}^{ct} = \mathcal{C} \, \frac{3\beta}{(4\pi)^2 v} \left(\frac{m^4}{\Lambda_Q^4}\right) H  F_{\alpha\beta} F^{\alpha\beta} \, ,
\label{ct_l}
\end{align}
as there is a unique (CP-even) dimension 6 effective-operator that satisfies the SM symmetries, that contributes to the corresponding process, and the same is valid for the $gg\to H$ process.

In the following, we shall revise the Compton scattering cross section and afterwards, calculate the new contributions to the $H\to \gamma\gamma$ and $gg\to H$ loop functions, which are, as we shall shortly demonstrate, the only relevant ones, as all the contributions to the remaining channels will turn out to be highly suppressed.

\section{Revised Compton Scattering}

For Compton scattering at high energies, we can safely work in the massless electron limit. The corrected Compton cross section (with respect to ref.~\cite{Delhom:2019wir}), including the standard QED contributions and the ones corresponding to the additional effective vertex from $\mathcal{L}^{\text{eff}}_{Q}$ (\ref{int_lagr}) is given by
\begin{align}
\label{compt}
\frac{d\sigma_{e\gamma\to e\gamma}}{d\Omega} &= \frac{1}{256 \pi^2 s} (5 + \cos^2\theta + 2 \cos\theta) \Bigg[ 
\left( \frac{3\beta}{4\Lambda_Q^4} \right)^2 s^4 (1+\cos\theta) \notag \\
& \qquad\qquad\qquad\qquad\qquad\qquad\qquad \qquad + \left(\frac{3\beta}{\Lambda_Q^4}\right) s^2 e^2 Q_e^2  + \frac{4 \, e^4 Q_e^4}{(1+\cos\theta)}  
\Bigg] \, ,
\end{align}
where $Q_e = -1$ is the charge of the electron and $e$ is the QED coupling constant. In the phenomenological analysis section (Section~\ref{phen}) we will re-analyse the corresponding experimental Compton scattering data from \cite{L3:2005uij} as in \cite{Delhom:2019wir} using the previous formula, and obtain bounds on the model parameters.

\section{${H \to \gamma\gamma}$ Decay Rate and $gg\to H$ Cross Section}
\label{hgg}

As mentioned previously, the three contributions to the ${H \to \gamma\gamma}$ effective vertex for on-shell photons and Higgs boson are shown in figure~\ref{eff_mu}, where the first diagram contains the $\bar{f}f\gamma\gamma$ effective vertex and the $H\bar{f}f$ Yukawa interaction, the second diagram corresponds to the $H\bar{f}f\gamma\gamma$ term from (\ref{int_lagr}) and, the last diagram corresponds to the counter-term. Its expression for one fermion in the loop takes the gauge-invariant form
\begin{align}
i \, \Gamma^{\mu\nu}_{H\gamma\gamma} = \, i (\eta^{\mu\nu} M_H^2 - 2 {q'}^\mu q^\nu) \Big(\Pi_F(m,M_H,\mu) + \Pi_{\epsilon}(m,\mu) \Big) \, , 
\label{effHgg}
\end{align}
where momentum conservation reads $p=q+q'$ and where we have neglected the terms that vanish when contracted with the photon polarization four-vectors $\epsilon^{\mu}_r (q)$ and 
$\epsilon^\nu_s(q')$. The $\Pi_\epsilon$ form factor contains an UV-pole, and its expression in the $\overline{\text{MS}}$ scheme is given by
\begin{align}
\Pi_{\epsilon}(m^2,\mu^2) = \mu^{2\epsilon} \frac{1}{2\hat{\epsilon}}  \left(\frac{m^4}{\Lambda_Q^4}\right) \frac{6\beta}{(4\pi)^2 v}  \, , 
\end{align}
where $\mu$ is the renormalization scale and $1/\hat{\epsilon} \equiv 1/\epsilon + \gamma_E - \ln(4\pi)$. This terms gets renormalized by the tree-level counter-term Lagrangian (\ref{ct_l}) as follows. We re-express the bare constant $\mathcal{C}$ appearing in the Lagrangian as the sum of the renormalized constant $\mathcal{C}_R(\mu)$ and an UV-divergent part that cancels the one that appears at one-loop in the $\Pi_\epsilon$ form factor i.e., 
\begin{align}
 \mathcal{C} = \mathcal{C}_R(\mu) + \delta_\mathcal{C}^\epsilon = \mathcal{C}_R(\mu) - \mu^{2\epsilon}\frac{1}{2\hat{\epsilon}} \, ,
\end{align}
therefore $\mathcal{C}_R(\mu)$ satisfies
\begin{align}
\mathcal{C}_R(\mu) = \mathcal{C}_R(\mu_0) + \ln(\mu/\mu_0) \, .
\label{crmu}
\end{align}
If we assume that this term is identically zero at some scale $\Lambda$, such that $\mathcal{C}_R(\Lambda) = 0$ then 
\begin{align}
\mathcal{C}_R(\mu) = \ln(\mu/\Lambda) \, .
\end{align}
In conclusion, after renormalizing we must make the substitution  
\begin{align}
 \Pi_{\epsilon}(m,\mu) \to  \Pi_{R}(m,\mu) = C_R(\mu) \left(\frac{m^4}{\Lambda_q^4}\right) \frac{6\beta}{(4\pi)^2 v} \, ,
\label{PIR}
\end{align}
where $C_R(\mu) = \ln (\mu/\Lambda)$. 

Going back to the expression (\ref{effHgg}), the finite form factor is given by
\begin{align}
\Pi_F(m,M_H,\mu) &= - \frac{3m}{2v}  \frac{1}{(4\pi)^2 \Lambda_Q^4} \bigg( m(2\alpha + \beta) (6m^2 - M_H^2) - \frac{4}{3}(4\alpha + \beta) \, m^3  \notag \\
& \qquad\qquad\qquad  -\beta m  \int_0^1 dx \Big[2m^2 + M_H^2  ( 6x(x-1) + 1 ) \Big] \ln \frac{a^2}{\mu^2} \bigg) \, ,
\label{PIF}
\end{align}
where $a^2 = m^2 + M_H^2x(x-1)$. After integrating in $x$, the final expression of the total form-factor renders finite and $\mu-$independent (however, it depends on the scale $\Lambda$) i.e.,
\begin{align}
\Pi(m,M_H) &\equiv \Pi_F(m,M_H,\mu) + \Pi_R(m,M_H,\mu) \notag 
\\
  &  =  \frac{1}{v \, (4\pi)^2} \left(\frac{m^4}{\Lambda_Q^4}\right) \left(  \frac{M_H^2}{m^2} (3\alpha + \beta) - (10\alpha + 7\beta) + 3\beta  \ln \frac{m^2}{\Lambda^2} \right)         \, .
\label{pif}
\end{align}

In our analysis there are two possible natural choices for $\Lambda$, that is, either $v$ or $\Lambda_Q$. Here, we shall vary the value of $\Lambda_Q$ in the interval $\Lambda_Q \in [v, 1000]$ GeV (where $v=246$ GeV) and present the results for the previous two choices, that is, $\Lambda = v$ and $\Lambda = \Lambda_Q$.

Including the SM fermionic and $W$ boson contributions, and extending the previous result to all SM fermions and their corresponding colours, the expression for the total decay width at tree level reads
\begin{align}
\Gamma(H\to\gamma\gamma) &= \frac{G_F  \alpha^2 M_H^3}{128 \, \pi^3} \bigg| \sum_f N_c^f  Q_f^2 \mathcal{F}(x_f) + \mathcal{G}(x_W) + \frac{4\pi v}{\alpha} \sum_f N_c^f  \Pi(m_f,M_H)  \bigg|^2  ,
\end{align}
where $x_f=4m_f^2/M_H^2$ (with $m_f$ the fermion mass), $Q_f$ and $N_c^f$ are the electric charge and the number of colors of the fermion $f$, and finally $x_W=4M_W^2/M_H^2$. The explicit expressions for the loop functions are, as usual, given by
\begin{align}
\mathcal{F}(x) = \frac{x}{2}[4+(x-1)f(x)] \, , \qquad \mathcal{G}(x) = -2 + 3x + \left( \frac{3}{2}x - \frac{3}{4}x^2 \right)f(x) \, ,
\end{align}
with
\begin{equation}
f(x)\; =\; \begin{cases} -4\arcsin^2(1/\sqrt{x})\, , \quad & x\geqslant1 \\[3pt] \Big[\ln\Big( \frac{1+\sqrt{1-x}}{1-\sqrt{1-x}}\Big)- i\pi \Big]^2\, , & x<1 \end{cases} \, .
\end{equation}

One should note, that the new contribution has a suppression factor ${m^4}/(4\pi{\Lambda_Q^4})$, but also an enhancement factor $(N_c/\alpha)$, with respect to the SM contributions. The overall effect is a suppression factor proportional to
\begin{align}
\left(\frac{m^4}{\Lambda_Q^4}\right) \frac{N_c}{4\pi\alpha} \; \sim \; 0.45 \, ,
\end{align}
for $\Lambda_Q = 500$ GeV and $m=m_t$. This result can be further enhanced with the $\alpha$ and $\beta$ parameters and obtain comparable contributions to the SM ones. 

Before continuing with our analysis further comments are required. In this study we are going to focus on the $H\to\gamma\gamma$ decay channel and also on the remaining decay modes ($WW,\, ZZ,\,\bar{b}b,\, \tau \tau$) but only the ones that are produced through the $gg$ fusion channel, which is affected by the same loop form factor as $H \to \gamma\gamma$. As for the neglected channels (production or decay), as they occur at the tree-level, the non-metricity induced interactions would only bring loop-suppressed corrections, which at this stage (given the current experimental uncertainties) can be safely neglected. For completeness, for a rough estimation of the order of magnitude of these corrections with an illustrative example, see Appendix~\ref{YUK}. 

The same form factors $\Pi_F$ and $\Pi_R$ also enter the gluon-fusion process. It can be trivially deduced that the $gg\to H$ cross section, including the non-metricity contributions, will be given by 
\begin{align}
\sigma(gg\to H) &= \frac{ M_H^2 \, \alpha_s^2}{1024 \, \pi \, v^2} \bigg| \sum_q  \mathcal{F}(x_q) + \frac{4\pi v}{\alpha_s} \sum_q \Pi(m_q,M_H)  \bigg|^2  \delta(s-M_H^2) \, .
\end{align}
Both the expression of the $gg$-fusion production cross section and the one for the $H\to \gamma\gamma$ will be needed in Section~\ref{phen-fits} when defining the LHC signal strengths.

\section{Phenomenology}
\label{phen}

This section will be dedicated to the phenomenological analysis of Compton scattering and the LHC Higgs signal strengths. From both we will obtain useful results with respect to the allowed regions on the parameter space of the model, namely on $\alpha$, $\beta$ and $\Lambda_Q$. As we shall see in the following, the bounds obtained from the LHC signal strengths are complementary to the ones obtained from Compton scattering, for a generic model where $\alpha$, $\beta$ and $\Lambda_Q$ are free parameters, however, the LHC constraints on the EiBI model will turn out to be the most stringent ones obtained up to date.

\subsection{Compton scattering}

\begin{figure}[!t]
\centering
\includegraphics[scale=0.55]{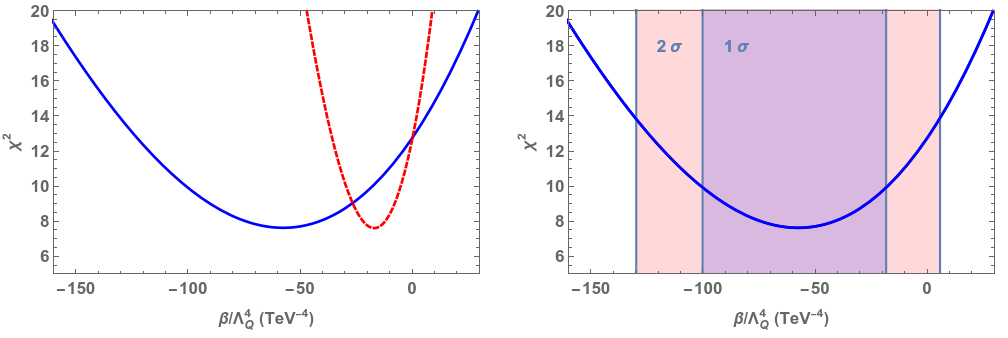} 
\caption{Left: $\chi^2$ function with the correct Compton scattering formula (blue continuous curve) and the previous results from \cite{Delhom:2019wir} (red-dashed curve). Right: $\chi^2$ function with the correct Compton scattering formula and the corresponding $1\sigma$ and $2\sigma$ bandwidths.}
\label{COMP}
\end{figure}

\begin{figure}[!t]
\centering
\includegraphics[scale=0.65]{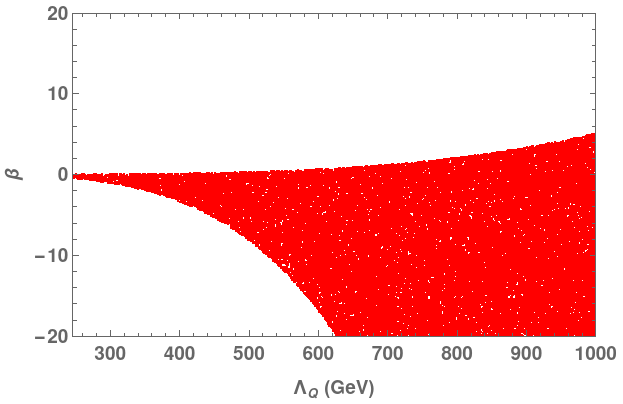}
\caption{Allowed $2\sigma$ region for the $(\Lambda_Q,\, \beta)$ parameter space obtained from the Compton scattering experimental data.}
\label{plot_lambda}
\end{figure}

As previously mentioned, we define a $\chi^2$ estimator including the Compton scattering experimental data \cite{L3:2005uij}, as in \cite{Delhom:2019wir}, for the Compton cross section (\ref{compt}). The results are shown in figure~\ref{COMP}. We can observe that the $\chi^2$ curve (blue) grows slower with $\beta/\Lambda_Q^4$ when moving away from the minimum when compared to the previous study \cite{Delhom:2019wir} (with the incorrect expression for the cross section, red-dashed curve). In the right panel, the $1\sigma$ and $2\sigma$ regions are also plotted using the correct Compton scattering formula. Depending on the sign of $\beta$, at the $2\sigma$ level we find 
\begin{align}
&|\beta|^{-4} \, \Lambda_Q > 0.29 \, \text{TeV} \qquad \text{with} \; \beta < 0 \, , \notag \\
&|\beta|^{-4} \, \Lambda_Q > 0.66 \, \text{TeV} \qquad  \text{with} \; \beta > 0 \, ,
\end{align}
which provides a less stringent bound for $\beta < 0$ and a more stringent bound for $\beta > 0$ than previously thought.\footnote{The previous results \cite{Delhom:2019wir} correspond to $|\beta|^{-4} \, \Lambda_Q > 0.39$ TeV for $\beta < 0 $ and, $|\beta|^{-4} \, \Lambda_Q > 0.61$ TeV for $\beta > 0$.} 
If, instead, we work in a two-parameter space $(\Lambda_Q, \,\beta)$, we obtain at $2\sigma$ level the allowed region shown in figure~\ref{plot_lambda}.

\subsection{Fits to the LHC data}

\label{phen-fits}

In order to be able to compare the model predictions with the experimentally measured signal strengths, we define the following ratios for the $H\to\gamma\gamma$ decay channel
\begin{align}
\mu_{\gamma\gamma}^X &\equiv
\frac{\sigma^X(pp\to H)\, \text{Br} (H \to \gamma\gamma)}{\sigma^X(pp\to H)_{\mathrm{SM}}\, \text{Br} (H\to \gamma\gamma)_{\mathrm{SM}}}\, ,
\end{align}
where $\sigma^X(pp\to H)$ stands for any specific production cross section. We also define the following ratios 
\begin{align}
\mu_{Y}^{gg} &\equiv
\frac{\sigma(gg\to H)\, \text{Br} (H \to Y)}{\sigma(gg\to H)_{\mathrm{SM}}\, \text{Br} (H\to Y)_{\mathrm{SM}}}\, ,
\end{align}
for the $H\to Y$ decays (where $Y$ stands for any measured final state), and where the production channel is gluon fusion.

In conclusion, given the considerations from the previous section, for this fit to the experimental LHC data, we will consider all production channels for the $H\to\gamma\gamma$ decay, and all decay channels for the $\sigma(gg\to H)$ production, as these are the only relevant signal strengths that can suffer sizeable modifications from the new parameters.

Introducing the quotients
\begin{align}
C_{\gamma\gamma} &= \frac{\Gamma(H\to \gamma\gamma)}{\Gamma(H\to \gamma\gamma)_{\text{SM}}} = \frac{\bigg| \sum_f N_c^f  Q_f^2 \mathcal{F}(x_f) + \mathcal{G}(x_W) + \dfrac{4\pi v}{\alpha} \sum_f N_c^f  \Pi(m_f,M_H)  \bigg|^2}{\bigg| \sum_f N_c^f  Q_f^2 \mathcal{F}(x_f) + \mathcal{G}(x_W)  \bigg|^2} \, ,
\notag\\
C_{gg} &= \frac{\sigma(gg\to H)}{\sigma(gg\to H)_{\text{SM}}} = \frac{\bigg| \sum_q  \mathcal{F}(x_q) + \dfrac{4\pi v}{\alpha_s} \sum_q \Pi(m_q,M_H)  \bigg|^2}{\bigg| \sum_q  \mathcal{F}(x_q) \bigg|^2} \, ,
\end{align}
and the $\rho_H$ function, that allows us to express the total decay of $H$, including the new interactions, in terms of the SM Higgs width as
\begin{align}
\Gamma_H = \rho_H \, \Gamma^{\mathrm{SM}}_H \, ,
\end{align}
the different signal strengths can be simply expressed as
\begin{align}
\mu_{\gamma\gamma}^{X} = C_{\gamma\gamma} \, \rho_H^{-1} \,, \qquad  \mu_{Y}^{gg} =  C_{gg} \, \rho_H^{-1} \, , \qquad \text{and}  \qquad \mu_{\gamma\gamma}^{gg} = C_{gg} \, C_{\gamma\gamma} \, \rho_H^{-1} \, ,
\end{align}
with $X=VBF, \, VH, \, \bar{t}tH$ and $Y=b\bar{b}, \, \tau^+\tau^-, \, VV^*$. For the following fit to the experimental data we shall use a $\chi^2$ estimator using the latest ATLAS and CMS experimental data \cite{ATLAS:2022fnp,ATLAS:2022yrq,ATLAS:2021tbi,ATLAS:2021pkb,ATLAS:2020wny,CERN-EP-2022-120,CMS:2022uhn, arXiv:2204.12957, CMS:2020gsy,CMS:2018uag,CMS:2018nsn, CMS:2020dvg,CMS:2017dib}.

\begin{figure}[!t]
\centering
\includegraphics[scale=0.55]{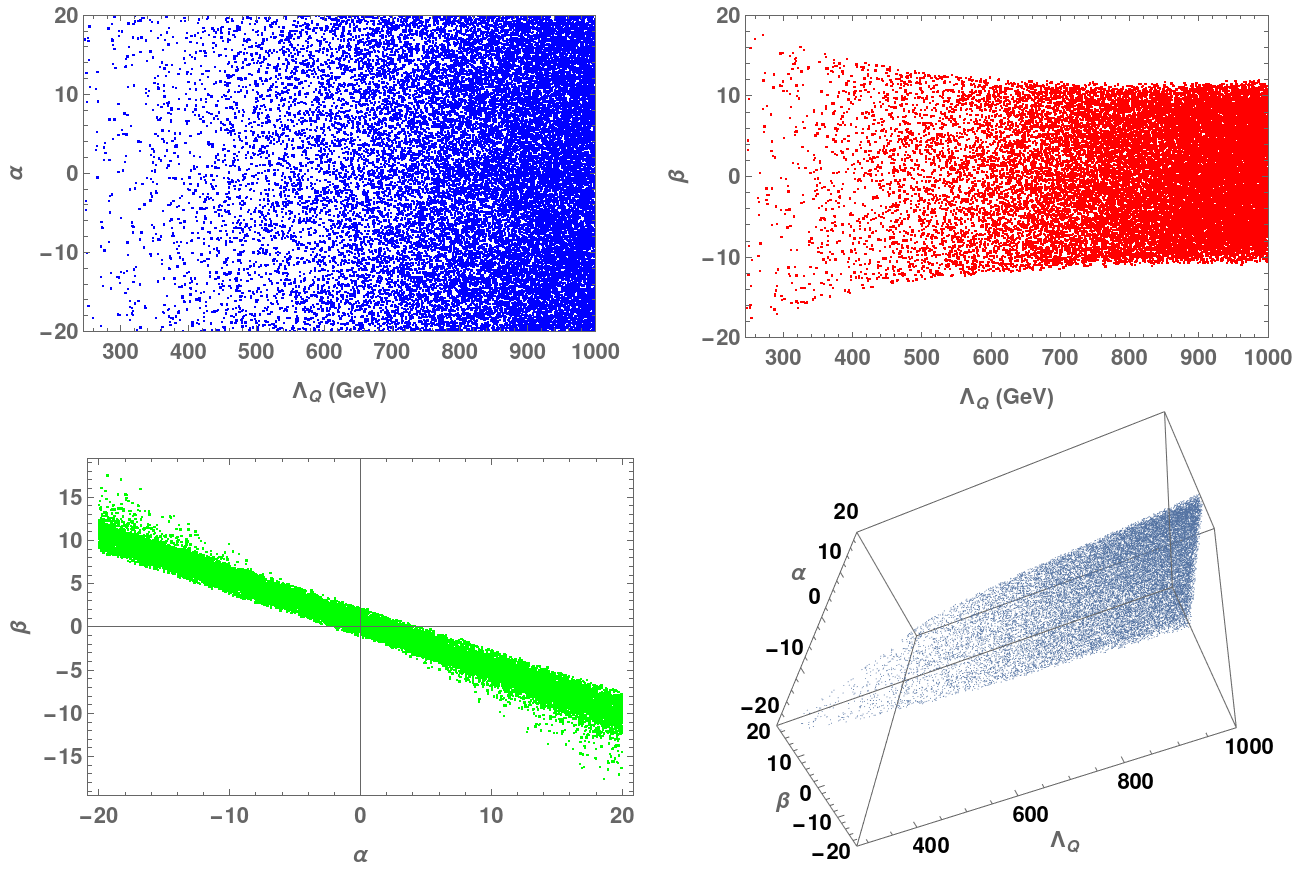} 
\caption{Allowed $2\sigma$ regions for different pair combinations of the model parameters (top-left/right and bottom-left), and for all three (bottom-right) for $\Lambda=\Lambda_Q$.}
\label{allowed1}
\end{figure}

If we choose the scale $\Lambda$ (at which the effective operator (\ref{crmu}) is zero, as explained in section~\ref{hgg}) as $\Lambda_Q$, i.e.,
$\Lambda=\Lambda_Q$, varying the parameters in the regions 
\begin{align}
\Lambda_Q \in [246, \, 1000 ] \, \text{GeV}\,, \qquad \alpha\in [-20,\, 20] \, , \qquad \beta\in [-20,\, 20] \, ,
\label{param} 
\end{align}
we obtain the $2\sigma$ allowed regions shown in figure~\ref{allowed1}. We observe that there is no correlation between $\alpha$ and $\Lambda_Q$ (top-left), however, we obtain an upper bound on $\beta$, roughly $|\beta|\lesssim 12$ depending on the value of $\Lambda_Q$ (top-right). On the other hand, we observe a very strong correlation between $\alpha$ and $\beta$ (bottom-left). This can be easily explained as follows. As the experimental data does not deviate significantly from the SM predictions, the new terms cannot bring large contributions therefore, the two terms must necessarily have opposite signs. Finally, we show the 3D allowed parameter space (bottom-right) which is, roughly, a 2D {\it surface}. 

Similar results can be obtained for $\Lambda=v$ for the same intervals given in (\ref{param}), except we obtain no upper bound for $|\beta|$. The main difference between the two choices of $\Lambda$ is given by the contribution of the term $\ln(m/\Lambda)$ in \eqref{pif} (with $m=m_t$, the dominant contribution). As this term only gets multiplied by $\beta$ (and not by a combination of $A \, \alpha + B \, \beta$, as all the other terms that contribute to the form factor) it translates into an effective upper bound on $|\beta|$. We can therefore conclude that, for values of $|\beta|>12$ the non-metricity contribution to the $\Pi(m,M_H)$ form factor \eqref{pif} becomes more sensitive to (the choice of) the scale $\Lambda$.

Once additional bounds are added, such as the previously analysed ones corresponding to Compton scattering, the parameter space, in both cases, gets drastically reduced. We shall specifically analyse these cases in the following section.  

Before moving on, let us shortly turn our attention to the specific EiBI and $f(R)$ models, as mentioned in the introduction. Fot the EiBI model, let us see what are the constraints that we can extract from the LHC data only. By setting $\alpha=0$ and $\beta=\pm 1$ we obtain from the $\chi^2$ fit,
\begin{align}
&\Lambda_Q > 835 \, \text{GeV} \qquad \text{with}   \qquad \Lambda = \Lambda_Q \, , \notag\\
&\Lambda_Q > 980 \, \text{GeV} \qquad \text{with}   \qquad \Lambda = v \, ,
\label{LHC1}
\end{align} 
for $\beta = -1$ and 
\begin{align}
&\Lambda_Q > 800 \, \text{GeV} \qquad \text{with}   \qquad \Lambda = \Lambda_Q \, , \notag\\
&\Lambda_Q > 690 \, \text{GeV} \qquad \text{with}   \qquad \Lambda = v \, ,
\label{LHC2}
\end{align} 
for $\beta  =1$. We can, therefore, conclude that the LHC constraints are yet, the most stringent ones obtained on the EiBI model.  

\begin{figure}[!t]
\centering
\includegraphics[scale=0.67]{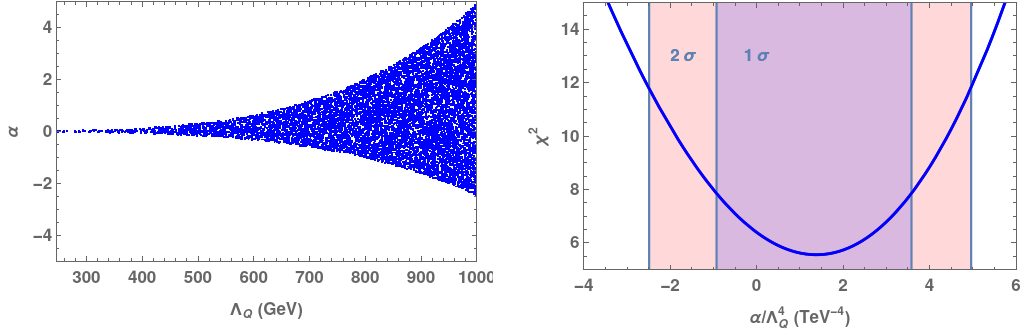} 
\caption{Left: allowed $2\sigma$ region for the $(\alpha,\, \Lambda_Q)$ parameter space. Right: $\chi^2$ function and the corresponding $1\sigma$ and $2\sigma$ bandwidths.}
\label{allowed22}
\end{figure}

As for the $f(R)$ models, they correspond to the $\beta = 0$ case and therefore, by taking a quick look at \eqref{pif}, one realizes that the form factor for this particular configuration does not depend on $\Lambda$. On the other hand, as they only depend on $\alpha$ (and $\Lambda_Q$), the constraints from the Compton Scattering analysis do not apply to these models, and so, the bounds on the corresponding $(\alpha,\Lambda_Q)$ parameter space will be determined by the LHC data only. Finally, from a $\chi^2$ fit we obtain the $2\sigma$ allowed region is shown in figure~\ref{allowed22} (left). If, instead, we choose only one parameter for the fit, namely $\alpha/\Lambda_Q^{-4}$ we obtain the results shown in figure~\ref{allowed22} (right). This roughly corresponds to 
\begin{align}
-2.5 < \frac{\alpha}{\Lambda_Q^4} < 5.0   \; (\text{TeV}^{-4})  \, ,
\end{align}
at $2\sigma$ level. These constraints can be reinterpreted in terms of any specific $f(R)$ model.

\subsection{Combined Constraints}

\begin{figure}[!t]
\centering
\includegraphics[scale=0.7]{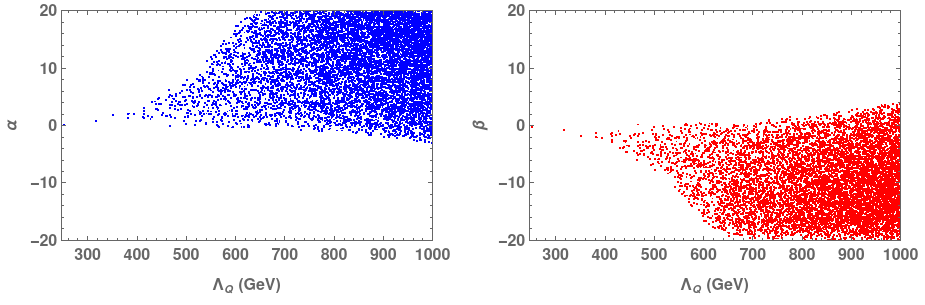} 
\caption{Allowed $2\sigma$ regions for $(\alpha,\, \Lambda_Q)$ and $(\beta,\, \Lambda_Q)$ parameter space for $\Lambda=v$.}
\label{allowedI}
\end{figure}

\begin{figure}[!t]
\centering
\includegraphics[scale=0.7]{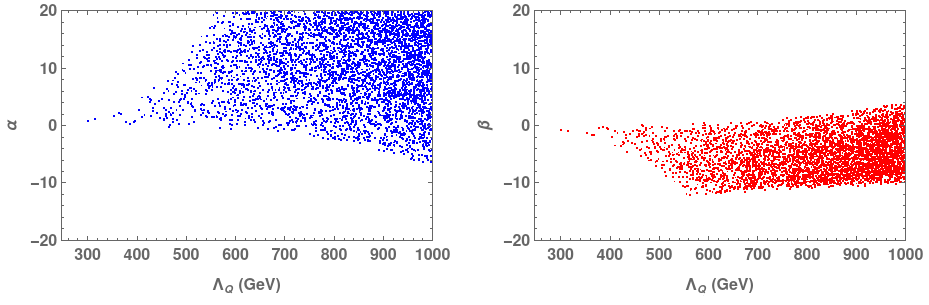} 
\caption{Allowed $2\sigma$ regions for $(\alpha,\, \Lambda_Q)$ and $(\beta,\, \Lambda_Q)$ parameter space for $\Lambda=v$.}
\label{allowedII}
\end{figure}

In the following, we shall combine the experimental constraints corresponding to the Compton scattering and the LHC data. The $2\sigma$ allowed regions are shown in figure~\ref{allowedI} for $\Lambda=\Lambda_Q$ and in figure~\ref{allowedII} for $\Lambda = v$. As expected, the Compton experimental data places stringent constraints on the $(\Lambda_Q,\, \beta)$ space, and the strong LHC constraint ($\alpha \simeq -\beta$) roughly translates into the fact that the allowed region in the $(\Lambda_Q, \, \beta)$ plane is a {\it mirror} image of the $(\Lambda_Q, \, \alpha)$ parameter space. For the $\Lambda=\Lambda_Q$ case, however, we can observe that there is a lower bound placed on $\beta$, which is given by the LHC data, and it is consistent with the results shown previously in figure~\ref{allowed1}.

If again, we turn our attention to the EiBI model, as the LHC data sets more stringent bounds on $\Lambda_Q$ for a one parameter fit, it should be clear that the previously obtained constraints \eqref{LHC1} and \eqref{LHC2} also apply for the combined fit. For the $f(R)$ models, as mentioned previously, the only data that can be applied for obtaining constraints on the parameter space are given by the LHC results only, therefore, the combined constraints would bring no new information.

\section{Conclusions}

In this manuscript, following the previous works \cite{Delhom:2019wir,Latorre:2017uve} we have studied a model originated by the form of the non-metricity tensor in RBG theories, by performing an expansion in inverse powers of the non-metricity scale $\Lambda_Q$. As a consequence we have a model with a reduced parameter space with terms that couple to the SM Lagrangian, thus it allows us to experimentally test non-Riemannian geometry associated effects at the LHC on Higgs-related high-energy experiments, which opens a new realm of possible future studies. Finally, we have obtained and discussed relevant bounds on its parameter space (the most stringent up to date). 

\begin{appendix}

\section{Interaction Lagrangian}
\label{INT_LAGR}

In this appendix we shall explicitly calculate the corresponding interaction Lagrangians for spin $J=1/2, \, 1,$ and $0$:
\begin{align}
\mathcal{L}_{(1/2)}^{\text{eff}}  &= \frac{i}{4\Lambda_Q^4} \left[\big[ (\mathcal{D}-1)\alpha + \beta \big] \, T^{(1)}\eta^{\mu\nu} - \beta \, T^{(1)\mu\nu} \right] \Big[ \bar{\psi} \gamma_\mu D_\nu \psi - 
(D_\nu \bar{\psi})\gamma_\mu \psi \Big] \notag \\
&  \qquad - \frac{m}{2\Lambda_Q^4} \bar{\psi} \psi (\mathcal{D}\alpha + \beta)T^{(1)} \, ,
\notag 
\\
\mathcal{L}_{(1)}^{\text{eff}}  &= \frac{1}{8\Lambda_Q^4} F_{\mu\alpha} F_{\nu\beta} \left[ \big[(\mathcal{D}-4) \alpha + \beta \big] \, T^{(1/2)}\eta^{\mu\nu} \eta^{\alpha\beta} - 4\beta \, \eta^{\mu\nu} \, T^{(1/2)\alpha\beta} \right]   \, ,\notag 
\\[1.5ex]
\mathcal{L}_{(0)}^{\text{eff}}  &= -\frac{m}{v}\frac{(\mathcal{D} \alpha + \beta)}{2\Lambda_Q^4} H \bar{\psi}\psi \, T^{(1)} \, .
\label{lag1}
\end{align}
The stress energy tensors for $J=1/2$ and $J=1$ can be obtained from (\ref{ini_lagr}) as usual i.e.,
\begin{align}
T^{(J)\mu\nu} = -\frac{2}{\sqrt{-g}}\frac{\partial \mathcal{L}^{(J)}}{\partial g^{\mu\nu}} \bigg|_{(\nabla_\mu, \, g_{\mu\nu}) \to (D_\mu,  \, \eta_{\mu\nu})  } \, ,
\end{align} 
Its explicit expressions for $J=1/2$ and $J=1$ read
\begin{align}
T^{(1/2)\mu\nu} &= \eta^{\mu\nu} \frac{i}{2}  \Big( \bar{\psi} \overleftrightarrow{\slashed{D}}\psi \Big) - \eta^{\mu\nu}  m \bar{\psi} \psi 
\notag \\ & \qquad  - \frac{i}{2}  \Big( \bar{\psi} (\gamma^\mu D^\nu + \gamma^\nu D^\mu) \psi - (D^\nu \bar{\psi} ) \gamma^\mu \psi - (D^\mu \bar{\psi} ) \gamma^\nu \psi \Big) \, ,
\notag \\
T^{(1)\mu\nu} &= \eta^{\mu\nu} \frac{1}{4} F_{\alpha\beta} F^{\alpha\beta} - \frac{1}{2}F^{\mu\alpha}F^{\nu}_{\;\; \alpha}  - \frac{1}{2}F^{\nu\alpha}F^{\mu}_{\;\; \alpha} \, .
\label{A1}
\end{align}
The corresponding traces in $\mathcal{D}$ dimensions are given by
\begin{align}
T^{(1/2)} &= T^{(1/2)\mu}_\mu = i \left({\mathcal{D}}/{2} - 1\right) \,  \bar{\psi} \overleftrightarrow{\slashed{D}}\psi - \mathcal{D} \, m \bar{\psi}\psi \, , 
\notag \\[1.5ex]
 T^{(1)} &= T^{(1)\mu}_\mu =  \left({\mathcal{D}}/{4}-1 \right) F_{\alpha\beta}F^{\alpha\beta} \, .
\label{A2}
\end{align}
Inserting expression (\ref{A1}) and (\ref{A2}) into (\ref{lag1}) and summing all three contributions we finally obtain the effective interaction Lagrangian $\lag^{\text{eff}}_Q = \mathcal{L}_{(1/2)}^{\text{eff}}  + \mathcal{L}_{(1)}^{\text{eff}} + \mathcal{L}_{(0)}^{\text{eff}}$, which reads
\begin{align}
 \lag^{\text{eff}}_Q &= \frac{f_1(\mathcal{D},\alpha,\beta)}{\Lambda_Q^4} \frac{i}{2} \left( \bar{\psi} \overleftrightarrow{\slashed{D}}\psi \right) F_{\alpha\beta} F^{\alpha\beta}  - \frac{f_2(\mathcal{D},\alpha,\beta)}{\Lambda_Q^4}\,  m \bar{\psi}  \psi \, F_{\alpha\beta} F^{\alpha\beta}  
 \notag \\ & \qquad + \frac{3 \, i \, \beta}{4\Lambda_Q^4} \, F^{\mu\alpha}F^{\nu}_{\;\;\alpha} \, \Big( \bar{\psi} \gamma_\mu D_\nu \psi - (D_\nu \bar{\psi})\gamma_\mu \psi \Big) -\frac{m}{v}\frac{f_3(\mathcal{D},\alpha,\beta)}{2\Lambda_Q^4}   H \bar{\psi}\psi \, F_{\alpha\beta}F^{\alpha\beta}  \, ,
\label{RR}
\end{align}
with $f_1$ and $f_2$ given by 
\begin{align}
f_1(\mathcal{D},\alpha,\beta)  &= \frac{1}{8} \Big[ \alpha \, (\mathcal{D} - 4)(2\mathcal{D} - 3) + \beta \, (2 \mathcal{D} - 11) \Big] = -\frac{3\beta}{8} + \frac{1}{8} (10\alpha + 4 \beta) \epsilon + \mathcal{O}(\epsilon^2) \, , \notag \\[1.5ex]
f_2(\mathcal{D},\alpha,\beta)  &= \frac{1}{4}  \left(\mathcal{D} - 4\right) \left( \mathcal{D} \, \alpha + \beta \right) = \frac{1}{2} (4\alpha + \beta) \epsilon + \mathcal{O} (\epsilon^2) \, ,
\notag \\[1.5ex]
f_3(\mathcal{D},\alpha,\beta)  &= (\mathcal{D} \alpha + \beta) \left({\mathcal{D}}/{4}-1 \right) = \frac{1}{2} (4\alpha + \beta) \epsilon  + \mathcal{O}(\epsilon^2)\, ,
\end{align}
where we considered, as mentioned previously $\mathcal{D}=4+2\epsilon$, expanded in $\epsilon$ and kept the $\mathcal{O}(\epsilon)$ terms.

Considering no additional interactions through $D_\mu$ i.e., taking $D_\mu \to \partial_\mu$, and using the equations of motion i.e., $(i/2)\bar{\psi} \overleftrightarrow{\slashed{\partial}}\psi - m \bar{\psi} \psi = 0$ we can further simplify the previous Lagrangian (\ref{RR}). We obtain
\begin{align}
 \lag^{\text{eff}}_{Q} &=   \frac{3 \, i \, \beta}{4\Lambda_Q^4} \, F^{\mu\alpha}F^{\nu}_{\;\;\alpha} \, \Big( \bar{\psi} \gamma_\mu \partial_\nu \psi - (\partial_\nu \bar{\psi})\gamma_\mu \psi \Big) - \left( \frac{\beta}{2} + \alpha \, \epsilon \right) \frac{3}{4 \Lambda_Q^4} m \bar{\psi}  \psi \, F_{\alpha\beta} F^{\alpha\beta}
\notag \\
 & -   \epsilon  \left( 4 \alpha  + \beta \right)  \frac{m}{v} \frac{1}{4 \Lambda_Q^4} H \bar{\psi}  \psi \, F_{\alpha\beta} F^{\alpha\beta}
 \, ,
\end{align}
which is the expression that we used in our analysis.

\section{Loop-induced Yukawa corrections}
\label{YUK}

\begin{figure}[!t]
\centering
\includegraphics[scale=0.4]{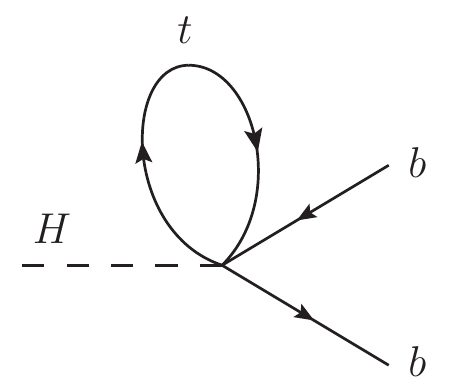}
\caption{One-loop correction to the $H\bar{b}b$ Yukawa coupling.}
\label{hbb_}
\end{figure}

In order to clarify why the non-metricity contributions are highly suppressed for processes that take place at tree-level let us take the following example. Consider the $\mathcal{L}_{(0)}^{\text{eff}}$ Lagrangian from (\ref{lag1}), written in a more generic form, as a sum over all SM fermions, and let us also consider that these terms additionally couple to the $T^{(1/2)}$ stress-energy tensor. In this case we obtain an additional interaction Lagrangian $\mathcal{L}_{(0)}^{' \text{eff}}$ given by the following expression
\begin{align}
\mathcal{L}_{(0)}^{' \text{eff}}  &= -\sum_j \frac{m_j}{v}\frac{(\mathcal{D} \alpha + \beta)}{2\Lambda_Q^4} H \bar{\psi}_j\psi_j \, T^{(1/2)}  \notag \\
&= -i\sum_j \frac{m_j}{v}\frac{(\mathcal{D} \alpha + \beta)}{2\Lambda_Q^4} H \bar{\psi}_j\psi_j  \sum_k \Big[ \left(\mathcal{D}/{2} - 1\right) \,  \bar{\psi}_k \overleftrightarrow{\slashed{D}}\psi_k - \mathcal{D} \, m_k \bar{\psi}_k\psi_k \Big] \, .
\end{align}
If we consider no additional interactions through $D_\mu$ and we apply the equations of motion, we obtain 
\begin{align}
\mathcal{L}_{(0)}^{' \text{eff}}  \,=  \, \sum_{j,k} \mathcal{C}_{jk} \frac{m_j  m_k}{\Lambda_Q^4 \, v}  H \bar{\psi}_j\psi_j\bar{\psi}_k\psi_k \, .
\end{align}
If we choose $\psi_j=b$ and $\psi_k = t$ we obtain the dominant loop correction to the $H\to \bar{b}b$ channel as shown in figure~\ref{hbb_}. Thus, the bottom Yukawa coupling receives a correction $\delta_b$ as follows
\begin{align}
\frac{m_b}{v} \big(1+ \delta_b \big) \qquad \text{with} \qquad \delta_b \sim \frac{1}{(4\pi)^2} \frac{1}{\Lambda_Q^4} \cdot (m_t^4,\; m_t^2 M_H^2) \sim \mathcal{O}(10^{-4}) \, 
\end{align}
for $\Lambda_Q = 500$ GeV, which is extremely small, and, as previously mentioned, can be safely discarded. Similar considerations can be made for the $H\to VV$ channels ($V=W,Z$) and for the different production channels, except of course, for the gluon fusion mechanism which will be taken into consideration.

\end{appendix}

\section*{Acknowledgements}

I would like to thank A. Pich for helpful comments on this manuscript.

%%%%%%%%%%%%%%%%%%%%%%%%%%%%%%%%%%%%%%%%%%%%%%%%%%%%%%%%%%%%%%%%%%%%%%%%%%%%%%%%%%%%%%%%%%%%%%%%%%%%%%%%%%%%%%%%%%%%%%%%%%
%%%%%%%%%%%%%%%%%%%%%%%%%%%%%%%%%%%%%%%%%%%%%%%%%%%%%%%%%%%%%%%%%%%%%%%%%%%%%%%%%%%%%%%%%%%%%%%%%%%%%%%%%%%%%%%%%%%%%%%%%%
%%%%%%%%%%%%%%%%%%%%%%%%%%%%%%%%%%%%%%%%%%%%%%%%%%%%%%%%%%%%%%%%%%%%%%%%%%%%%%%%%%%%%%%%%%%%%%%%%%%%%%%%%%%%%%%%%%%%%%%%%%

\bibliographystyle{JHEP}
\bibliography{cite}

\end{document}